\documentclass[5p]{elsarticle}

\usepackage{amsmath}
\usepackage[amssymb]{SIunits}

\usepackage{booktabs}

\usepackage[nodots,nocompress]{numcompress}

\usepackage{doi}
\usepackage{hyperref}

\begin{document}

\title{Metal hydride material requirements for automotive hydrogen storage systems}

\author[utrc]{Jos{\'e} Miguel Pasini}
\ead{pasinijm@utrc.utc.com}

\author[srnl]{Claudio Corgnale}

\author[utrc]{Bart A. van Hassel}

\author[srnl]{Theodore Motyka}

\author[gm]{Sudarshan Kumar}

\author[pnnl]{Kevin L. Simmons}

\address[utrc]{United Technologies Research Center, 411 Silver Lane, East Hartford, CT 06108, USA}

\address[srnl]{Savannah River National Laboratory, Savannah River Site, Aiken, SC 29808, USA}

\address[gm]{Chemical Sciences and Materials Systems Lab, General Motors Global R\&D, Warren, MI 48090, USA}

\address[pnnl]{Pacific Northwest National Laboratory, 902 Battelle Boulevard, P.O. Box 999, Richland, WA 99352, USA}

\begin{abstract}
The United States Department of Energy (DOE) has published a progression of technical targets to be satisfied by on-board rechargeable hydrogen storage systems in light-duty vehicles. By combining simplified storage system and vehicle models with interpolated data from metal hydride databases, we obtain material-level requirements for metal hydrides that can be assembled into systems that satisfy the DOE targets for 2017. We assume minimal balance-of-plant components for systems with and without a hydrogen combustion loop for supplemental heating. Tank weight and volume are driven by the stringent requirements for refueling time. The resulting requirements suggest that, at least for this specific application, no current on-board rechargeable metal hydride satisfies these requirements.
\end{abstract}


\maketitle

\section{Introduction}

Energy-efficient cars emitting zero greenhouse gases: this ultimate goal makes fuel cell vehicles running on hydrogen a very attractive concept. Furthermore, light-duty vehicles represent a large fraction of the commercial vehicle market, so it is important to focus on this class of vehicles. Due to the special volume and weight constraints for this sector of the market, as well as the expectations generated from the performance of currently existing vehicles, hydrogen-powered vehicles must satisfy certain performance requirements if they are to compete with light-duty vehicles based on other technologies.

Starting with performance demands on the vehicle, the United States Department of Energy (DOE) has produced a series of performance targets for the hydrogen storage system itself~\cite{DOETargets2009}. The targets are divided into phases, to guide the pace of technology improvement by research and development teams. The three phases correspond to 2010, 2017, and Ultimate Full Fleet targets.\footnote{The phases used to be 2010, 2015, and Ultimate Full Fleet. The 2015 phase has been changed to 2017.} The ultimate targets are designed to describe vehicles that would be competitive with other light-duty vehicles in the market.

Even though the DOE hydrogen storage targets have been modified over recent years, the basis for the targets remains the same: to develop a vehicle operating on hydrogen whose performance is not compromised when compared to today's current vehicles.  Currently, there are over 20~targets listed by DOE that deal with the characteristics of the onboard hydrogen storage system as it relates to vehicle performance, energy efficiency, safety, cost as well as the system's overall size and weight.  While all the targets are important, some, such as volumetric and gravimetric density and cost, are among the more difficult targets to meet for many hydrogen-based systems.

In this paper we use simplified storage system and vehicle models, combined with minimal balance-of-plant components, to obtain the material-level requirements for metal hydrides to satisfy the 2017 system-level targets. Because we are dealing with hypothetical materials, some additional assumptions are needed in order to complete the analysis, and we turn to metal hydride databases to anchor these assumptions. Sec.~\ref{sec:approach} details the approach and Sec.~\ref{sec:additional_assumptions} makes explicit some additional assumptions. In Sec.~\ref{sec:results} these assumptions come together to result in material requirements.

\section{Hydrogen storage materials and baseline metal hydride systems}

Because hydrogen is a lightweight gas at normal conditions, compressing it to high pressures (350 and 700~bar) and liquefying it at extremely low temperatures (20K) have become common storage methods.  While both of these ``physical'' storage methods are being applied to onboard hydrogen vehicles today, neither can meet all of the DOE targets and many scientists and researchers are actively pursuing other options.

One of these options that has and continues to be evaluated is storage of hydrogen on solid materials. This can include adsorption on high-surface materials by low energy molecular bonding (physisorption) or by absorption into materials with stronger chemical bonding (chemisorption). Like physical storage methods, solid storage methods also have deficiencies for automotive applications.  Physisorption systems, while not requiring liquid hydrogen temperatures, still require cryogenic temperatures near to that of liquid nitrogen (77K) to store enough hydrogen for today's vehicle requirements. Chemisorption systems can be divided into two material classes: reversible and non-reversible systems.  For automobile applications, the non-reversible systems are systems based on materials that can not be readily recharged with gaseous hydrogen at a fueling station. These systems, that we often refer to as chemical hydride systems, require off-board recharging where more complex processing conditions can be carried out to successfully rehydrogenate the material~\cite{Aardahl2009, Devarakonda2012}.  Another class of materials, reversible materials, typically can be recharged with hydrogen under conditions typically found at some of today's gaseous or liquid refueling hydrogen stations.  These systems are often referred to as metal hydride sytems.  Table~\ref{tab:mat_matrix} shows some of the candidate adsorbent, chemical hydride and metal hydride materials that are currently being evaluated for DOE by the Hydrogen Storage Engineering Center of Excellence (HSECoE)~\cite{HSECoE2012}.

\begin{table}[tb]
\centerline{%
\includegraphics[width=93mm]{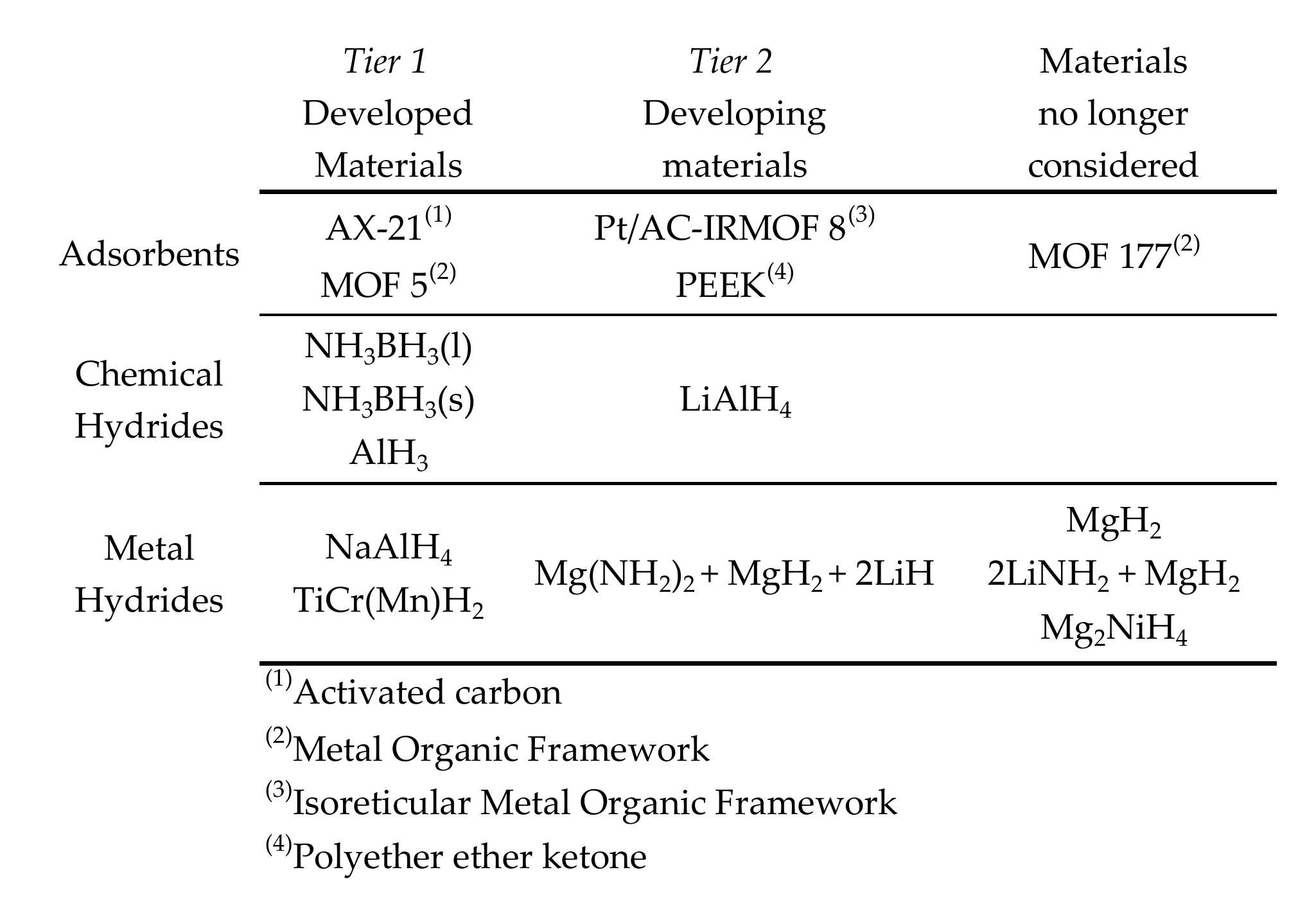}}
\caption{Hydrogen storage materials considered by the HSECoE, currently and in the past.}
\label{tab:mat_matrix}
\end{table}

\subsection{Metal hydride materials}

Over the past 30 years several hydrogen storage systems based on reversible metal hydrides have been evaluated for vehicle applications~\cite{Sandrock2003}. Most of these have involved either pure metals (like Mg) or, more commonly, intermetallic alloys (like LaNi$_5$, TiCrMn, and FeTi) as metal hydrides.  While hydrogen storage systems based on these metal hydride materials were often able to reversibly store and deliver hydrogen suitably for several industrial vehicle applications~\cite{Jacobs1998,Fuchs2001,Heung2001}, most of these systems were considered as being much too heavy for today's commercial vehicle market.

In the late 1990s some hydride materials, like NaAlH$_4$, which were widely considered to be non-reversible, were shown to be reversible under reasonable operating conditions with the addition of certain additives~\cite{Bogdanovic2000}.  Since then, several demonstration projects~\cite{Mosher2007, NaRanong2009, Johnson2011, Utz2011, BellostavonColbe2012} have evaluated NaAlH$_4$ as a possible reversible metal hydride for vehicle applications.  While improvements in the overall weight of a storage system appears possible using higher capacity sodium alanate material, the slower absorption and release rates for this material, coupled with higher heats of reaction or enthalpy, has resulted in little overall improvement of these systems for vehicle applications.

\subsection{Baseline systems}

Two metal hydride systems have received most attention as potential on-board hydrogen storage systems for light-duty vehicles. The first system is based on sodium alanate (NaAlH$_4$) while the second system is based on a high-pressure metal hydride (Ti$_{1.1}$CrMn). These two metal hydrides, because of significantly different operating conditions and kinetics of hydrogen absorption reactions, require substantially different system configurations.

Hydrogen absorption/desorption in the sodium alanate system is governed by the following reactions:
\begin{align}
\mathrm{NaAlH}_4 &\longleftrightarrow \frac{1}{3} \mathrm{Na_3 AlH_6 + \frac{2}{3} Al + H_2 } \label{eq:sah_rxn1} \\
\mathrm{ \frac{1}{3} Na_3 AlH_6 } &\longleftrightarrow \mathrm{ NaH + \frac{1}{3} Al + \frac{1}{2} H_2 } \label{eq:sah_rxn2}
\end{align}
The average enthalpy of this two-step reaction is approximately 41~kJ/mol-H$_2$ \cite{Bogdanovic2002}. At 3~bar, decomposition of the tetrahydride phase (NaAlH$_4$) starts at about 55$\degree$C, while decomposition of the hexahydride phase (Na$_3$AlH$_6$) starts at about 130$\degree$C.  The theoretical capacity of the sodium alanate system is 5.6~wt\%, but the practical capacity for an on-board hydrogen storage system utilizing sodium alanate is lower because of the slow kinetics of the absorption reactions and the need for additives to enhance the kinetics and thermal conductivity of sodium alanate. Because of the high enthalpy of reactions \eqref{eq:sah_rxn1} and~\eqref{eq:sah_rxn2}, thermal management of this system is quite challenging during refueling. In addition, the system requires combustion of a part of the hydrogen stored on-board to provide the heat of desorption whenever power is generated by conventional polymer electrolyte fuel cells (PEMFCs) operating below $\sim$360~K. However, this issue could be significantly mitigated for NaAlH$_4$ storage systems in applications where the waste heat comes from PEMFCs operating in the range 430--470~K, as described in~\cite{Pfeifer2009}.

On the other hand, the metal hydride Ti$_{1.1}$CrMn absorbs hydrogen reversibly at temperatures suitable for use with fuel cell vehicles~\cite{Kojima2006,Mori2009, Komiya2010}. In addition, the enthalpy of hydrogen absorption is 22~kJ/mol \cite{Sakintuna2007}, or about half of the enthalpy for the sodium alanate system. However, these materials absorb significant amounts of hydrogen only at high pressures and, even at such high pressures, these materials have low hydrogen capacity (1.9--2.0~wt\%).  The heat generated during refueling causes the bed temperature to rise resulting in lower absorption rates. Thus to ensure fast absorption, cooling the metal hydride during refueling is essential.  Because of lower reaction temperature and relatively lower heat of desorption, a major advantage of this system is that the fuel cell waste heat can be utilized, and there is no need to burn hydrogen during the desorption process. Thus this system offers higher on-board efficiency compared to the sodium alanate system. An additional advantage of this system is its high volumetric density. 

Both sodium alanate and Ti$_{1.1}$CrMn systems have been investigated by various authors using different heat exchanger configurations.   Ahluwalia \cite{Ahluwalia2007} considered a shell-and-tube heat exchanger with the metal hydride packed in the shell and the coolant flowing in tubes, while Johnson et al.\ \cite{Johnson2011} considered the metal hydride packed in the tubes and the heat transfer fluid flowing inside the shell. For the Ti$_{1.1}$CrMn system, Visaria et al.\ \cite{Visaria2011,Visaria2011a,Visaria2012} considered shell-and-tube heat exchangers with both straight and coiled cooling tubes. As part of the DOE HSECoE, we have investigated systems for both sodium alanate and Ti$_{1.1}$CrMn. The sodium alanate system has been described in Refs.~\cite{Raju2011} and~\cite{Kumar2012}, while the Ti$_{1.1}$CrMn system has been described in some detail in~\cite{Raju2010}.  For the sodium alanate system, we used a two-dimensional COMSOL\textsuperscript{\textregistered}
 model for evaluating refueling dynamics as well as heat transfer coefficients for the system level model. System level performance of these storage systems during driving conditions has been evaluated using a simulation model developed in the MATLAB/Simulink\textsuperscript{\textregistered} platform.

\begin{figure}[tb]%
\centerline{%
\includegraphics[width=90mm]{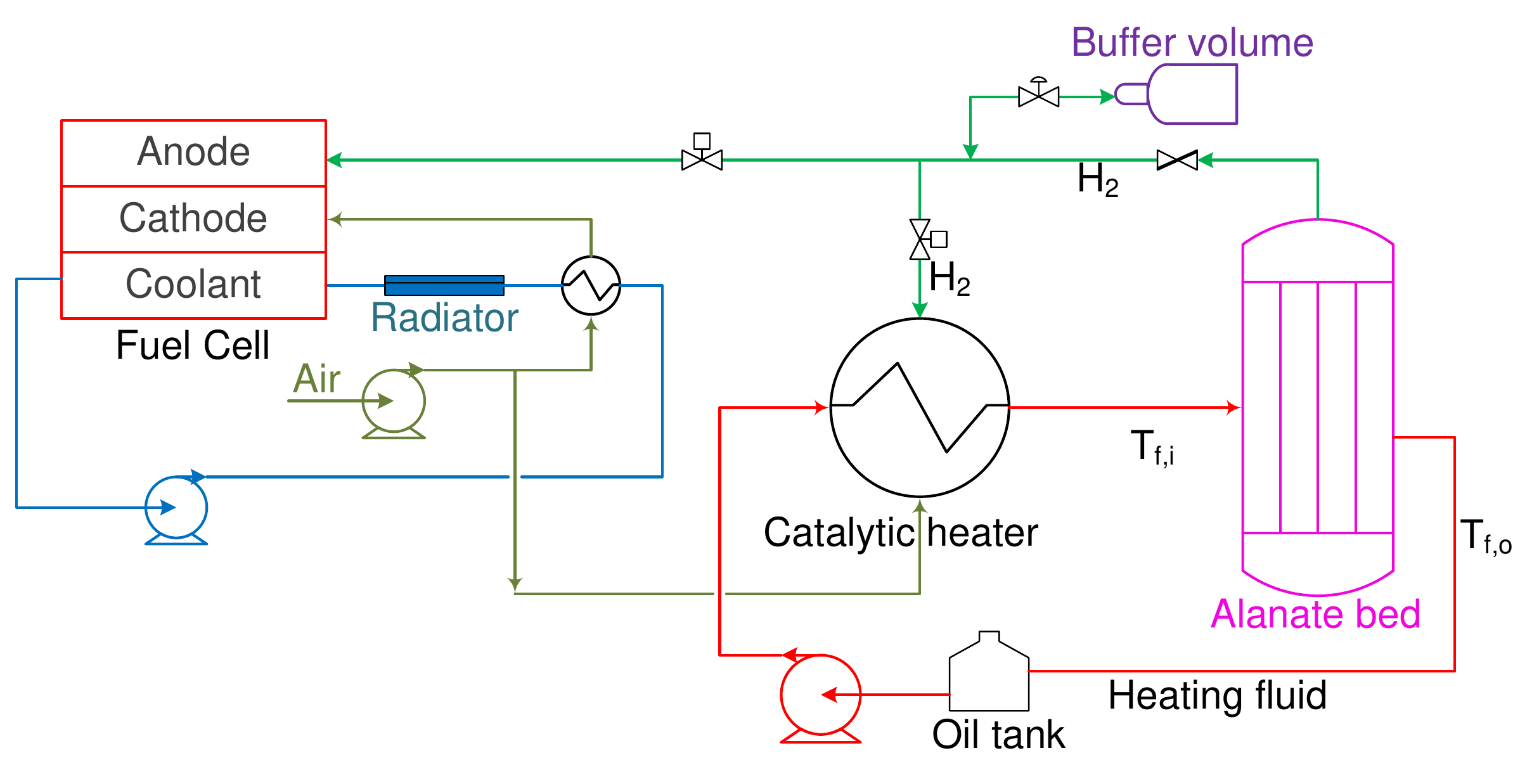}}%
\caption{Schematic diagram of the sodium alanate system.}%
\label{gm_sah_system}%
\end{figure}

For both the sodium alanate and the Ti$_{1.1}$CrMn system, the heat generated during refueling must be removed.  Fig.~\ref{gm_sah_system} shows the schematic for the sodium alanate system. The heat exchanger configuration adopted for this system is the shell-and-tube type heat exchanger with metal hydride in the shell. For this system, the gravimetric and volumetric capacities are reported to be approximately 0.012~kg-H$_2$/kg and 0.015~kg-H$_2$/L~\cite{Kumar2012}.

Similarly, for the Ti$_{1.1}$CrMn system, a similar shell-and-tube heat exchanger system was used.  Since the desorption reaction takes place at temperatures lower than 85$\degree$C, there is no need to burn hydrogen stored on-board. For this system, the heat of desorption can be supplied by the radiator fluid. The heat exchanger employed here is of the same design as that for the sodium alanate system. Because of the low absorption capacity of the Ti$_{1.1}$CrMn material, the system gravimetric and volumetric densities are less than 0.0145~kg-H$_2$/kg and 0.05~kg-H$_2$/L, respectively~\cite{Raju2010}.

It is clear that these two representative metal hydride systems fall significantly short of the DOE 2017 performance targets for hydrogen storage systems (0.055~kg-H$_2$/kg and 0.04~kg-H$_2$/L).  Therefore, we need to consider the properties that a metal hydride must possess in order to meet the DOE goals.

\section{Approach}

\label{sec:approach}

\begin{figure}[tb]%
\centerline{%
\includegraphics[width=90mm]{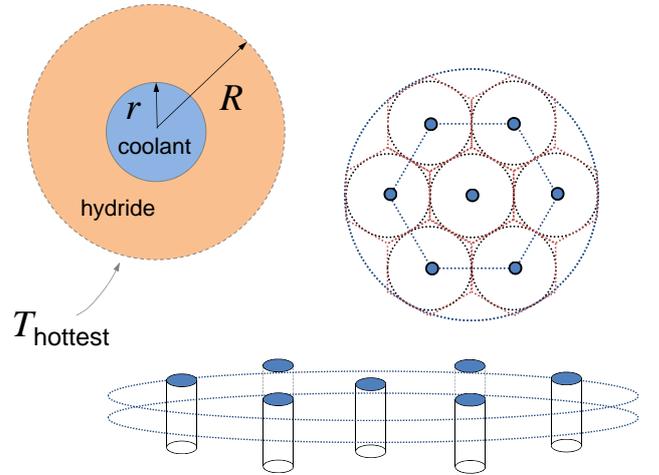}}%
\caption{Geometry of the unit cell.}%
\label{unit_cell}%
\end{figure}

The goal of this study is to understand what material-level requirements on metal hydrides are capable of yielding a system that satisfies the DOE 2017 targets. A critical component of the system is the tank containing the hydride and additional elements to ensure the system can reject the heat released during the fast refueling time. A higher enthalpy material, for example, will require that cooling tubes be placed closer to each other (see Fig.~\ref{unit_cell}), resulting in a tank with more space taken by non-storage material.

At the highest level, the approach used here is simple: we choose a value of the enthalpy and capacity of the pure material and obtain, in the end, the weight and volume of the system with minimal balance-of-plant components that can be refueled in the 2017 target time (corresponding to 1.5~kg-H$_2$/min~\cite{DOETargets2009}). The system must also be able to deliver 5.6~kg of hydrogen to the fuel cell under standardized driving conditions~\cite{Pasini2012}.

Here we describe our procedure. We start by choosing a value for enthalpy and capacity of the pure material. This capacity is then lowered by assuming 10\% (by weight) of additional inert material for enhancing thermal conductivity (see Sec.~\ref{sec:thermal_conductivity}). We then assume that only 85\% of this maximum capacity is achieved during the short refueling time. By estimating the amount of hydrogen that needs to be combusted by the system to effect the dehydrogenation, we calculate the mass of hydride contained in the pressure vessel.

Next, we estimate the spacing between cooling tubes by using the Acceptability Envelope method~\cite{Corgnale2012}, which is summarized in Sec.~\ref{sec:acceptability_envelope}. With the cooling tube spacing calculated and the total mass of hydride, we obtain the internal volume required for the pressure vessel. This internal volume, together with the vessel's pressure rating, defines the vessel wall thickness, and therefore its external volume and also its additional contribution to the system weight (Sec.~\ref{sec:pressure_vessel}). This concludes the tank design.

Once we have a tank design based on the acceptability envelope, we implement that design in our integrated vehicle simulation framework (Sec.~\ref{sec:framework}). Thus we verify that the system is indeed capable of providing 5.6~kg of hydrogen to the fuel cell when driven from full to empty under a standardized driving scenario.

\subsection{Acceptability envelope}
\label{sec:acceptability_envelope}

In this section we briefly summarize the Acceptability Envelope method~\cite{Corgnale2012}.

The acceptability envelope is a scoping tool which identifies the range of chemical, physical and geometrical parameters for a coupled media and storage vessel system, allowing it to reach determined performance targets. The simplified model which underpins the analysis, based on steady state refueling conditions, allows the energy balance equation to be expressed as follows: 
\[
\nabla^2 T = -\frac{\dot{Q}}{k},
\]
where $k$ is the thermal conductivity in the bed and $\dot{Q}$ is the heating rate per unit volume (positive during exothermic hydrogen absorption). This heating rate is given by
\[
\dot{Q} = \left(\frac{-\Delta H}{W_{\text{H}_2}}\right) \left(\frac{\Delta m_{\text{H}_2}}{\Delta t}\right) \left(\frac{M_\text{hydride}}{\rho_\text{bed}}\right)^{-1},
\]
where $\Delta H$ is the overall enthalpy per mole of H$_2$ absorbed, $W_{\text{H}_2}$ is the molar mass of H$_2$, $\Delta m_{\text{H}_2}$ is the target mass of hydrogen to add to the material, $\Delta t$ is the target refueling time, $M_\text{hydride}$ is the mass of hydride (including any additional inert material), and $\rho_\text{bed}$ is the density of the hydride bed, including inert material and porosity. The three terms in parentheses are, respectively, the heat released per mass of H$_2$ absorbed, the refueling rate, and the volume of hydride in the tank including voids.

The assumed geometry is shown in Fig.~\ref{unit_cell}, with the tank approximated as multiple unit cells with symmetry conditions at the outer radius. This implies that the boundary conditions assumed to solve the energy balance equation, for a single unit, see a fixed temperature at radius equal to~$r$ with the maximum temperature achieved at the symmetry border located at~$R$. Consequently the temperature rise between the coolant tube and the outer edge of the unit cell is
\begin{equation}
\Delta T = \frac{1}{8} \frac{\dot{Q}}{k} (R^2 - r^2) f(x),
\label{eq:temperature_rise}
\end{equation}
where $R$ is the cell outer radius, $r$ is the coolant tube outer radius, and
\begin{eqnarray*}
x &=& \frac{R^2}{r^2} - 1 \\
f(x) &=& 2 \left(1 + \frac{1}{x} \right) \ln (1 + x) - 2.
\end{eqnarray*}

By fixing the refueling rate (from the targets), the material properties, $\Delta T$ (assumed in this paper to be 45{\degree}C, within the range used in~\cite{Corgnale2012}), and the coolant tube radius~$r$, we can solve for~$R$, thus determining the amount of additional metal and tube space and their contribution to the weight and volume. This, in turn, determines the total tank internal volume. In Sec.~\ref{sec:pressure_vessel} we consider the effects that variations in internal volume have on the pressure vessel itself.

\subsection{Enthalpy and equilibrium pressure}
\label{sec:enthalpy}

The equilibrium pressure of a hydride at different temperatures plays a central role in the design of the hydrogen storage system. It affects tank wall thickness, as well as whether the fuel cell waste heat can be used for releasing hydrogen. In this work we must deal with hypothetical materials, yet their assumed properties must be based on available information from existing materials. With this goal in mind, we turn to a database of metal hydride properties~\cite{HydrogenMaterialsDatabase2011}. For the specific case of equilibrium pressure, we use the enthalpy as a variable and we fit the entropy to obtain the pressure through the van 't Hoff equation
\[
P_{\mathrm{H}_2}  \; [\mathrm{atm}] = \exp\bigl[(\Delta H - T \Delta S)/RT \bigr].
\]

\begin{figure}[tb]%
\centerline{%
\includegraphics[width=90mm]{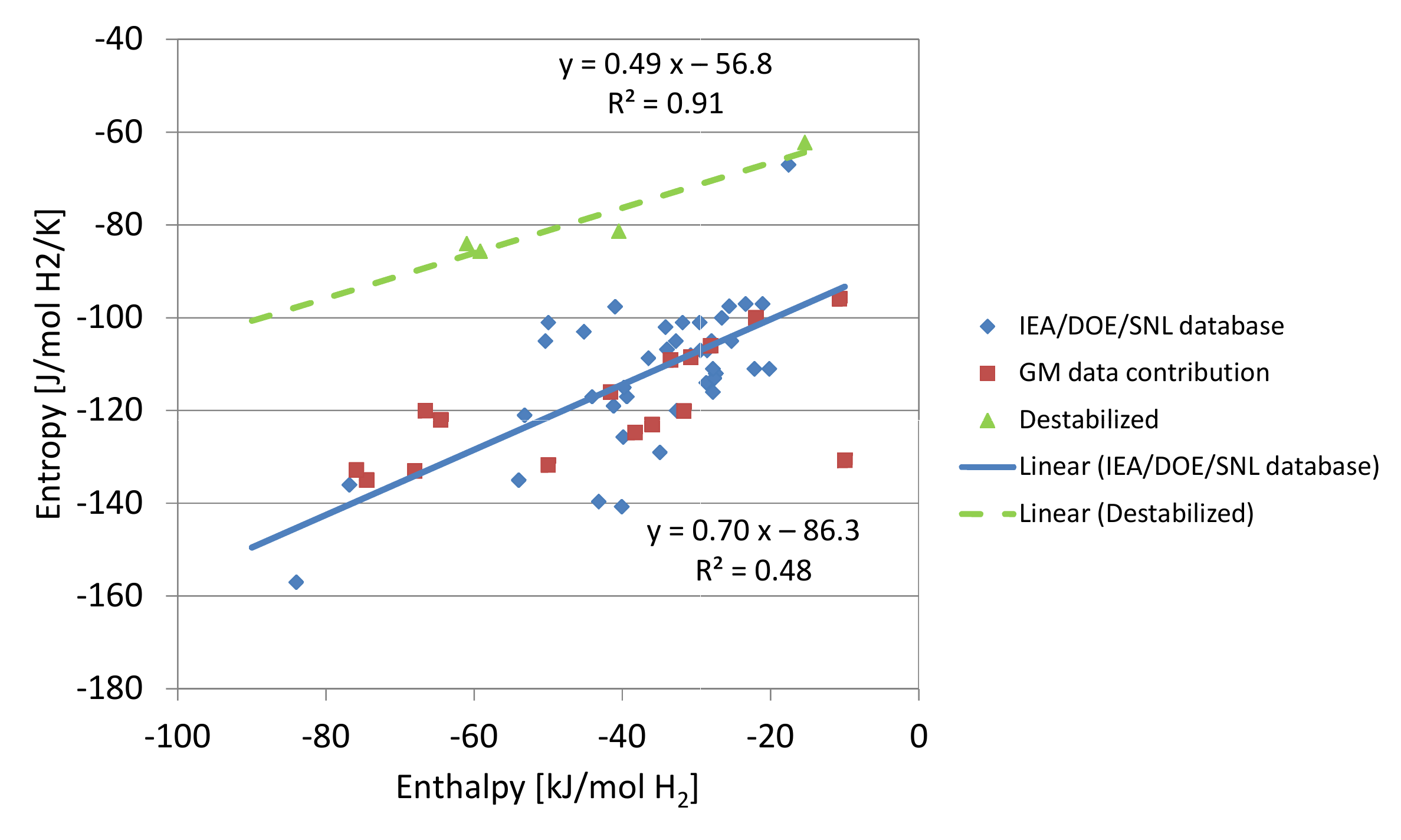}}%
\caption{Entropy as a function of enthalpy per mole of H$_2$ absorbed. Solid line: fit to the IEA/DOE/SNL database. Dashed line: fit to destabilized lithium borohydrides.}%
\label{entropy_vs_enthalpy}%
\end{figure}

\begin{figure}[tb]%
\centerline{%
\includegraphics[width=90mm]{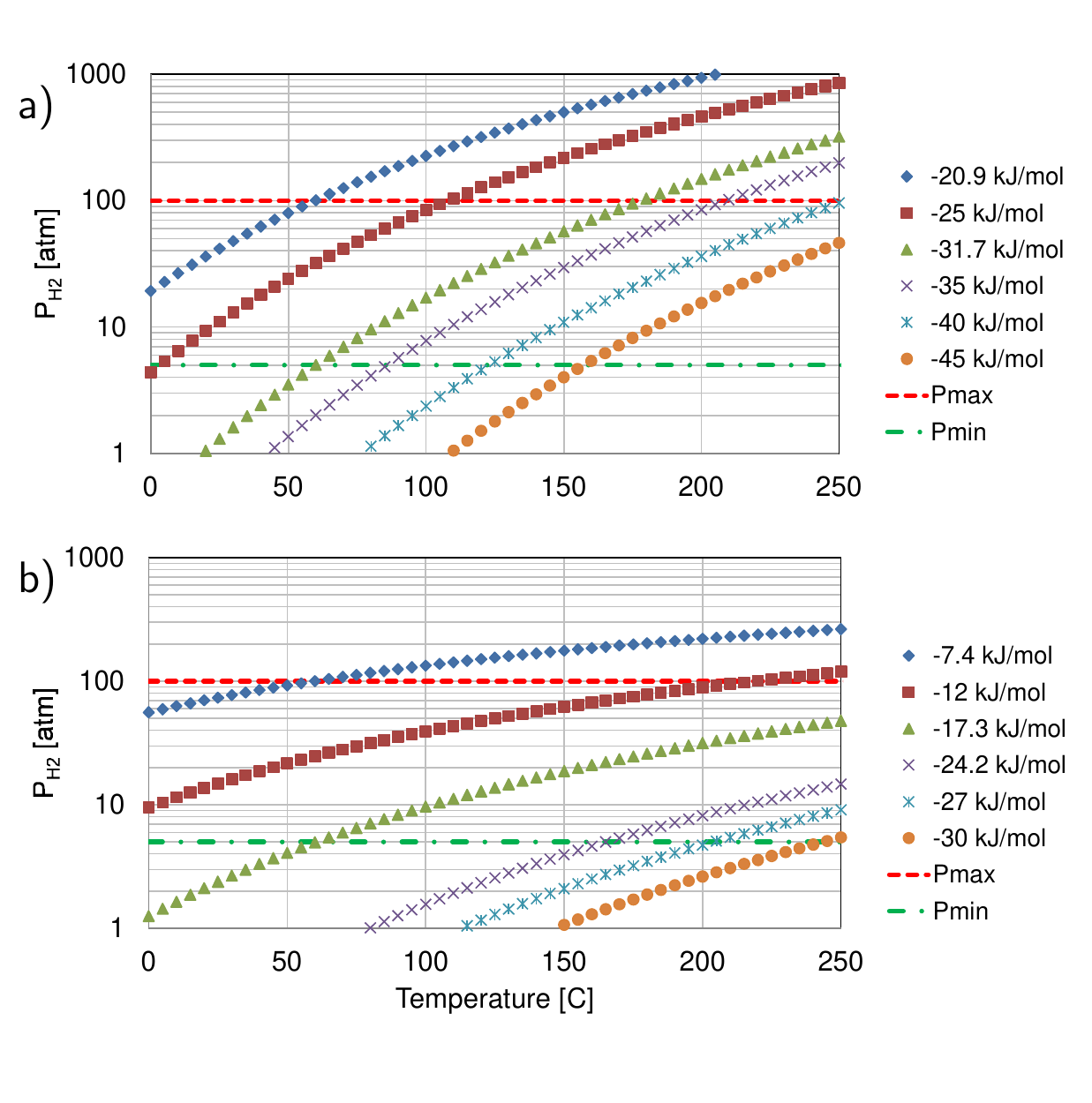}}%
\caption{Equilibrium pressure as a function of temperature for (a) complex metal hydrides and (b) destabilized systems.}%
\label{equilibrium_pressure}%
\end{figure}

Fig.~\ref{entropy_vs_enthalpy} shows the entropy as a function of enthalpy. We see a clear distinction between two populations: complex hydrides form a cluster with consistently higher entropy (more negative) than destabilized lithium borohydrides. In those destabilized systems, the composition is altered with an additive to increase pressure and decrease operating temperature at the cost of reduced gravimetric capacity and usually no improvement in kinetics~\cite{Vajo2007}. Due to this separation in equilibrium pressure, we use two separate fits to make two sets of predictions for material requirements. In Fig.~\ref{equilibrium_pressure} we present the equilibrium pressure resulting from the two fits in Fig.~\ref{entropy_vs_enthalpy}.

\subsection{Density and porosity}
\label{sec:density}

\begin{figure}[tb]%
\centerline{%
\includegraphics[width=90mm]{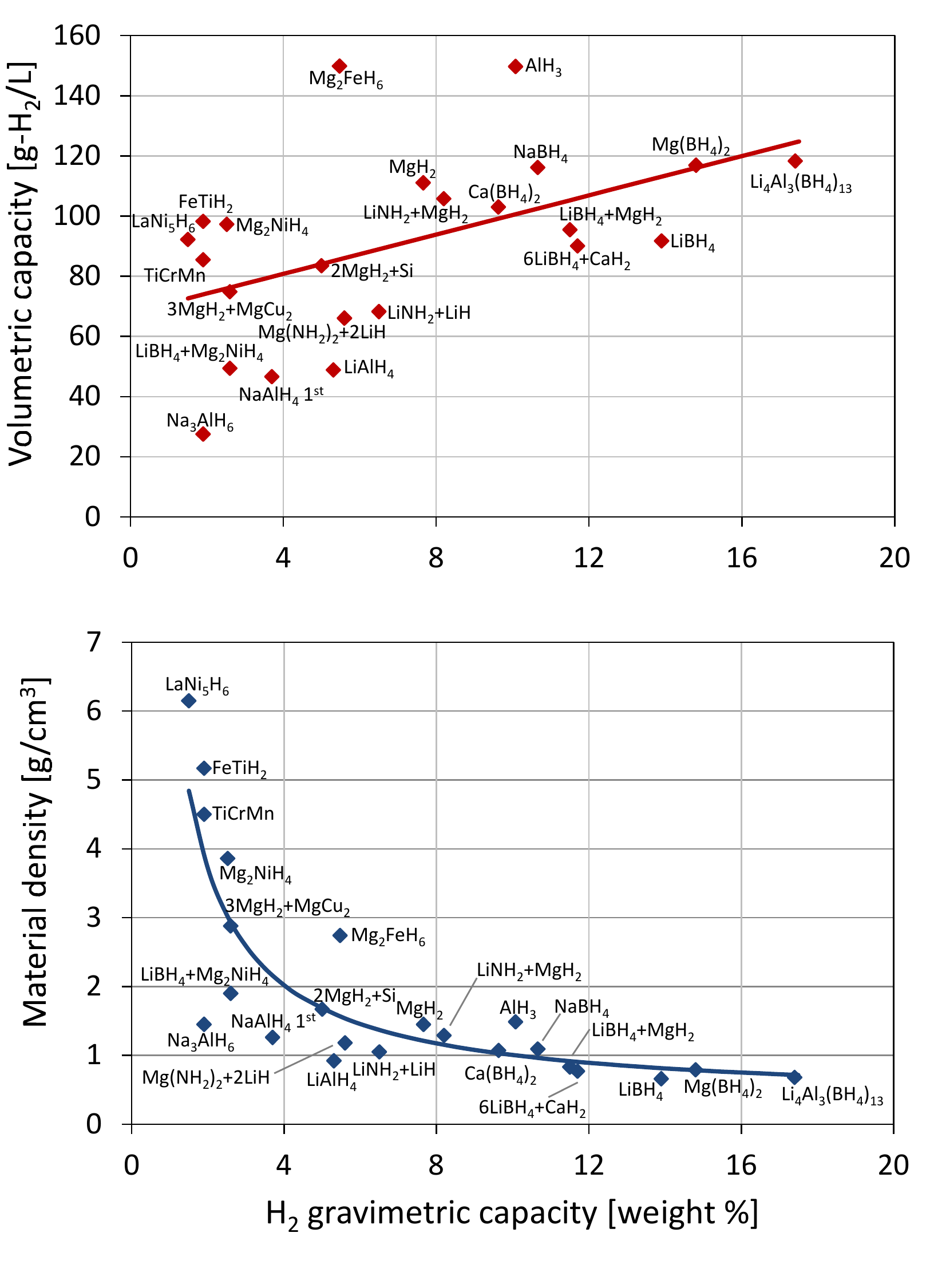}}%
\caption{Volumetric capacity (top) and density (bottom) as a function of the pure material gravimetric capacity. The solid bottom curve corresponds to the linear fit of the top graph.}%
\label{density_volumetric_capacity}%
\end{figure}

The stringent volume constraints in a light-duty vehicle are represented by the DOE target for system-level volumetric storage capacity. The material density therefore plays an important role in the viability of the system as a whole. Fig.~\ref{density_volumetric_capacity} shows the crystal density of the pure material in the hydrided state and the associated volumetric storage capacity as a function of its gravimetric capacity. We note a clear trend for the density: as the gravimetric capacity increases, the density decreases.

Given the gravimetric capacity, the volumetric capacity and density are related, so a fit for one data set implies a fit for the other. We chose to do a linear fit of the volumetric capacity and calculate from it the following fit for the crystal density:
\begin{equation}
\rho_\text{crystal} \left[\frac{\text{kg}}{\text{m}^3}\right] = 326.25 + \frac{67.74}{w_\text{crystal}},
\label{eq:density_correlation}
\end{equation}
where $w_\text{crystal}$ is the theoretical weight fraction of the material.\footnote{Note that the scatter for the volumetric capacity is high and the correlation is weak. This is reflected in the linear fit.}

In Ref.~\cite{vanHassel2012} NaAlH$_4$ powder is compacted into pellets, reducing the volume by~42\%. After 15 absorption-desorption cycles, however, the volume of the NaAlH$_4$ pellets had increased by $\sim$25\% from their as-prepared sizes. In this study we assume that the material has been compacted to a fixed 30\% porosity. Note that, since the density fit is for the hydrided state (in which the material takes up more volume), this is the minimum porosity of the material, which would occur after refueling.

\subsection{Pressure vessel}
\label{sec:pressure_vessel}

At this stage, we have the internal volume and weight of the pressure vessel. To obtain the wall thickness and concomitant additional weight and volume, we assume refueling to 100~bar and then use the simple fit from Ref.~\cite{Pasini2012}. There the authors derived, based on data provided by Lincoln Composites for a Type~IV tank rated for refueling to 100~bar, a linear fit for the additional weight and volume due to the vessel walls. In this study the choice of a Type~IV tank is driven by the need to have minimal weight. As discussed in~\cite{Pasini2012}, for each operating temperature, one would need to ensure that the polymer liner material typical of Type~IV tanks can be used.

\subsection{Integrated vehicle simulation framework}
\label{sec:framework}

At this stage in the approach, we have the pressure vessel design, with both its internal and total weight and volume. To validate some of the assumptions made so far, such as the amount of additional hydrogen to combust while driving, we turn to simulation at the vehicle level. The goal is to ensure that the system with this tank design and minimal balance-of-plant components (see Sec.~\ref{sec:bop}) can deliver 5.6~kg of hydrogen to the fuel cell under standardized transient conditions.

Ref.~\cite{Pasini2012} describes in detail the vehicle simulation framework used here. It is designed to allow changing the storage system easily without changing any assumptions for the rest of the vehicle and the fuel cell system. 

The simulation framework also includes five Test Cases, corresponding to different drive cycles and ambient conditions. In this study we use Test Case 1, the ``ambient drive cycle.'' It alternates between the 7.5-mile, 22.8-minute Urban Dynamometer Driving Schedule (UDDS) and the 10.26-mile, 12.75-minute Highway Fuel Economy Driving Schedule (HWFET). This alternation is repeated indefinitely until the storage system fails to deliver the flow rate requested by the fuel cell. The UDDS represents low speeds and stop-and-go urban traffic, while the HWFET represents free-flow traffic at highway speeds.

\section{Additional assumptions}

\label{sec:additional_assumptions}

\subsection{Thermal conductivity}
\label{sec:thermal_conductivity}

Apart from the enthalpy and material density discussed in Secs. \ref{sec:enthalpy} and~\ref{sec:density}, another key quantity that appears in the Acceptability Envelope approach is the thermal conductivity.

Since rejecting heat while refueling is a limiting process, enhancing the thermal conductivity of metal hydrides is critical. Ref.~\cite{vanHassel2012} shows how the thermal conductivity of NaAlH$_4$ can be enhanced by adding material, such as aluminum or graphite. In particular, they show that by adding Expanded Natural Graphite (ENG) worms, the increase is nearly optimal as compared with the parallel model of thermal conduction in a mixture.

For this study we assume that 10\% by weight of ENG worms has been added to the hydride and that this addition results in the value $k = 9\;\mathrm{W} \mathrm{m}^{-1} \mathrm{K}^{-1}$.

\subsection{Material kinetics}

The weight fraction of the material $w$ is assumed to follow the single-step kinetics
\begin{eqnarray*}
\frac{dw}{dt} &=& - \text{sgn} (z) z^\chi w_\text{full} A e^{-E_a/RT} \left|\ln \left(\frac{P}{P_\text{sat}(T)}\right) \right| \\
z &=&   \frac{w}{w_\text{full}} - x_\text{sat},
\end{eqnarray*}
where sgn is the sign function and $x_\text{sat} = 1$ if $P > P_\text{sat}$ and zero if not. $w_\text{full}$ is the weight fraction of H$_2$ in the hydride when fully loaded. This is not the theoretical material capacity, because at this stage we already assume that 10 wt\% ENG worms has been added to enhance the thermal conductivity. We chose $\chi=2$ to ensure that the asymptotic approach to the full state is polynomial and not exponential, which would be overly optimistic. For the activation energy we chose $E_a = 110 \; \text{kJ/mol}$, based on previous fits to NaAlH$_4$ data~\cite{vanHassel2012}.

\subsection{Balance-of-plant components}
\label{sec:bop}

The HSECoE created a library of balance-of-plant (BOP) components for use by system architects and designers. It includes components that are currently available off-the-shelf and was built by surveying previous system design work, literature, specific component design, and by discussing with component suppliers.

The system architects and designers working within the HSECoE draw from the library to build a bill of material that documents the quantity, mass, volume, and component requirements from the system diagrams while documenting the source of key components.  This enables them to track specific components in the bill of material of the designed system. The bill of material is then used to develop cost models.

\begin{figure}[tb]%
\centerline{%
\includegraphics[angle=180,width=90mm]{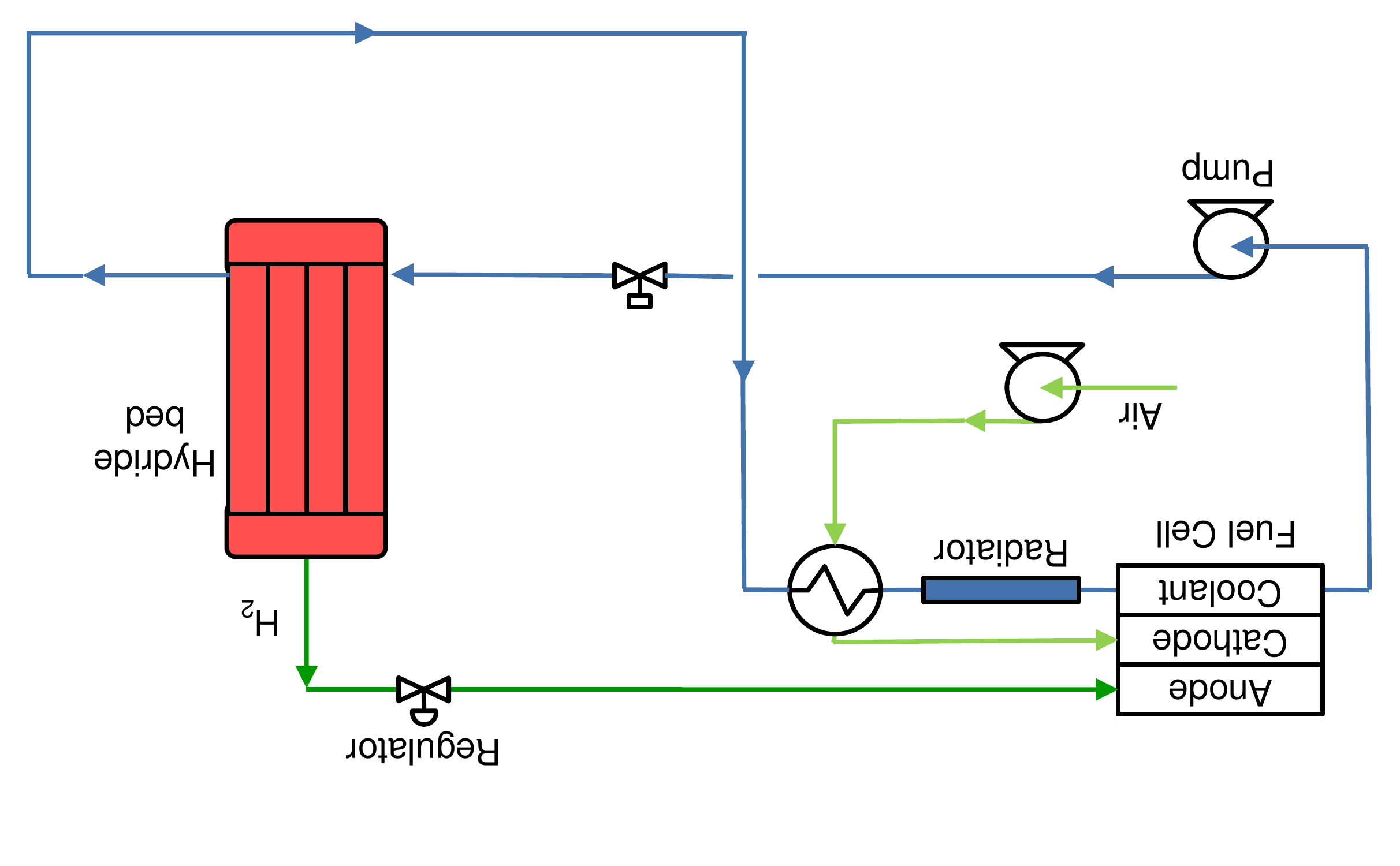}}%
\caption{Diagram for a minimal system that does not require combustion for the dehydrogenation process.}%
\label{minimal_system_no_combustion}%
\end{figure}

Fig.~\ref{minimal_system_no_combustion} shows an ideal system that does not require combustion for the dehydrogenation process. This is the case for materials with high equilibrium pressure, such as TiCrMn~\cite{Raju2010}. In such systems, the coolant leaving the fuel cell stack is hot enough that it can drive hydrogen release in the tank.

Such a simple system would require only minimal BOP components. The top part of table~\ref{tab:bop} shows the minimal balance-of-plant components for a system that doesn't need a combustor loop. This part of the table is based on Ref.~\cite{Hua2011}, where TIAX and Argonne National Laboratory detail the BOP components for a 350-bar compressed gas system.

\begin{table}[tb]
\centerline{%
\includegraphics[width=85mm]{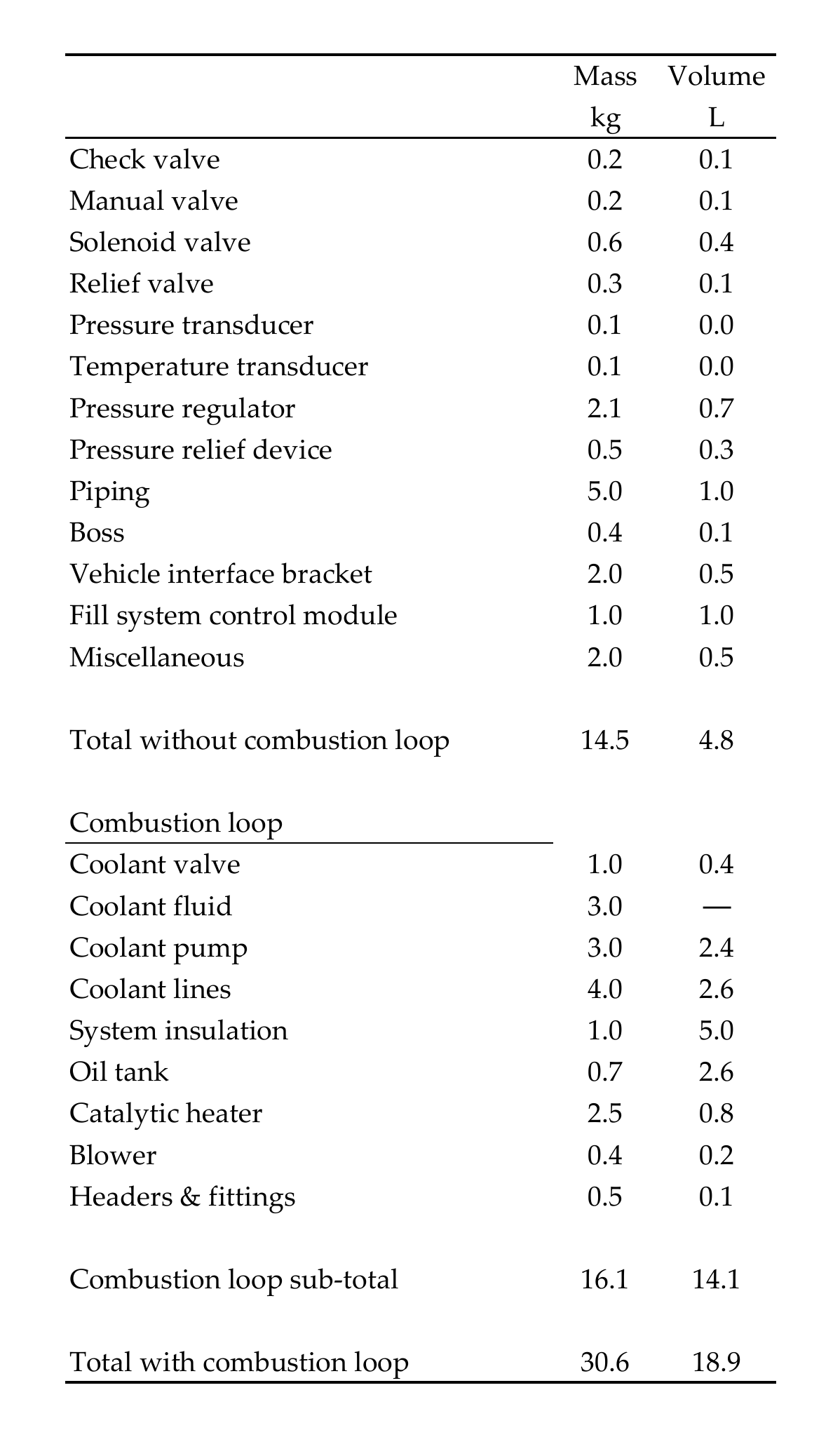}}
\caption{Minimal balance-of-plant components for a system without a combustion loop (top) and with a combustor (bottom).}
\label{tab:bop}
\end{table}

\begin{figure}[tb]%
\centerline{%
\includegraphics[angle=180,width=90mm]{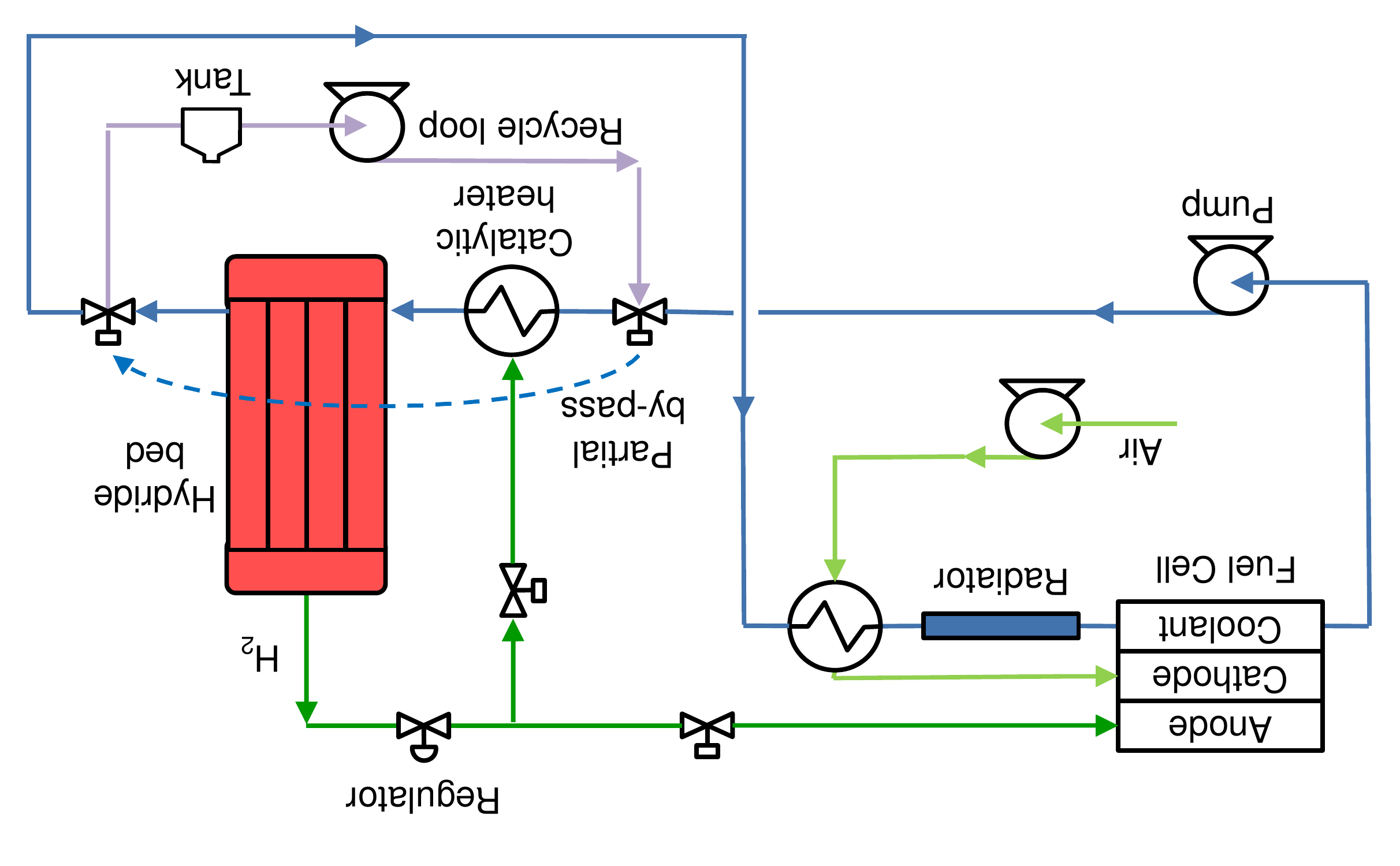}}%
\caption{Diagram for a minimal system that requires combustion for dehydrogenation but can make partial use of the the fuel cell's waste heat.}%
\label{minimal_system_with_combustion}%
\end{figure}

The bottom section of table~\ref{tab:bop} lists the additional components required by a combustion loop (Fig.~\ref{minimal_system_with_combustion}), as would be needed by a material with lower equilibrium pressure. The 2.5~kg, 0.8~L catalytic heater  corresponds to a 8~kW microchannel combustor and heat exchanger, sized by Oregon State University using the model described in Ref.~\cite{Ghazvini2011}.

For higher enthalpy materials we need a more powerful heater, such as the 30~kW combustor detailed in Ref.~\cite{Johnson2012}.

\section{Results}
\label{sec:results}

\subsection{Amount of hydrogen combusted}
\label{sec:h2_combusted}

\begin{figure}[tb]%
\centerline{%
\includegraphics[angle=270,width=90mm]{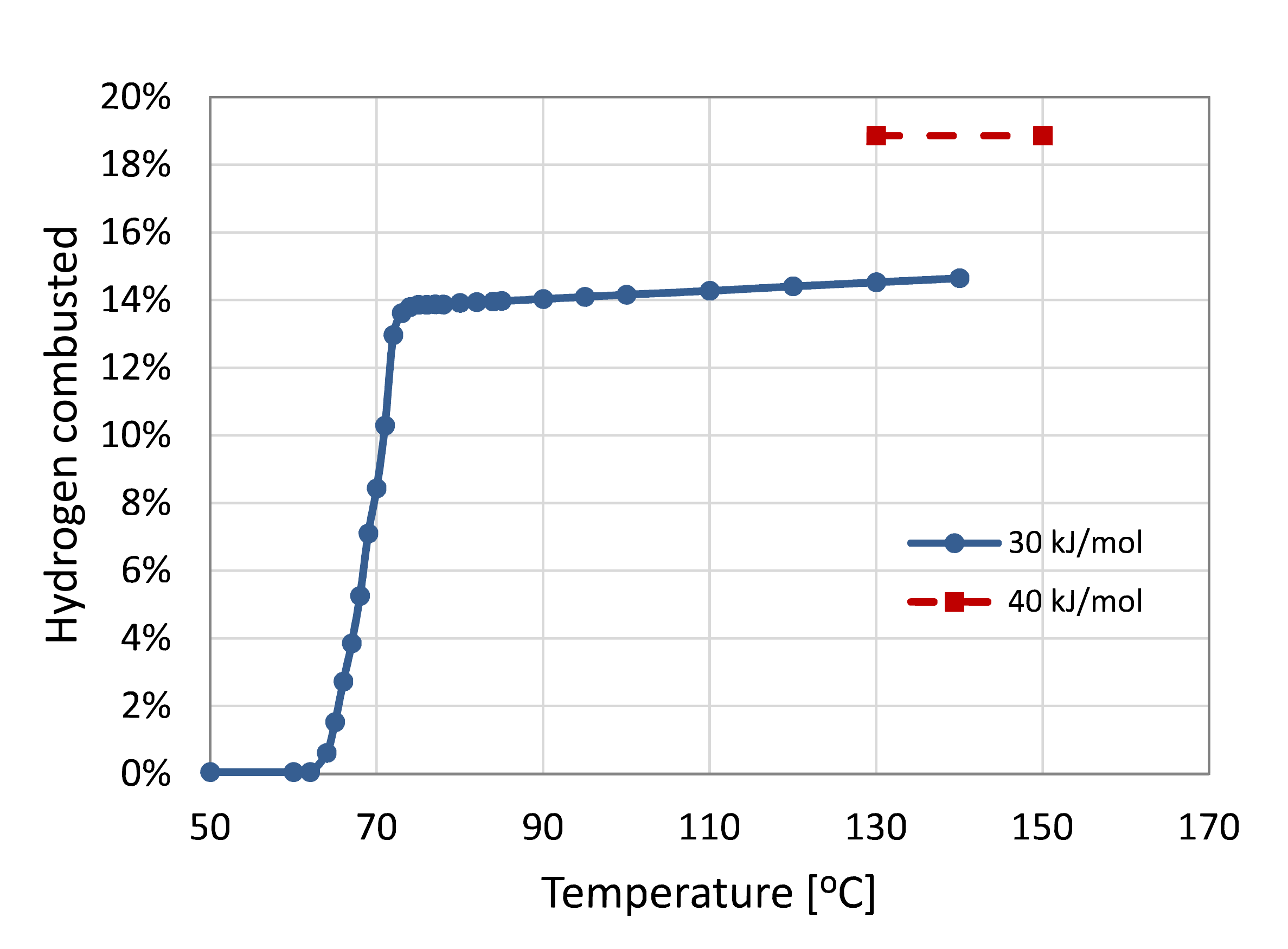}}%
\caption{Fraction of the total amount of hydrogen released by the tank that needs to be combusted as a function of the tank's temperature set point when the vehicle is running Test Case~1.}%
\label{h2_combusted}%
\end{figure}

Even a system that in principle can use the fuel cell waste heat may need to operate at higher temperature to compensate for slow kinetics. Fig.~\ref{h2_combusted} shows the result of simulating a vehicle under Test Case~1. As the set point is raised from 60{\degree}C to 72{\degree}C, the 30~kJ/mol system goes from no combustion to using 14\% of the hydrogen to maintain the tank conditions.

The higher enthalpy material shown in Fig.~\ref{h2_combusted}, due to its equilibrium pressure, has to operate at higher temperature in order to release H$_2$ at high enough pressure. The minimum delivery pressure is 5~bar. If we want the equilibrium pressure in the tank to be at least 6~bar, the operating temperature would have to be at least 130{\degree}C (see Fig.~\ref{equilibrium_pressure}).

\subsection{Tank weight and volume}

\begin{figure}[tb]%
\centerline{%
\includegraphics[width=90mm]{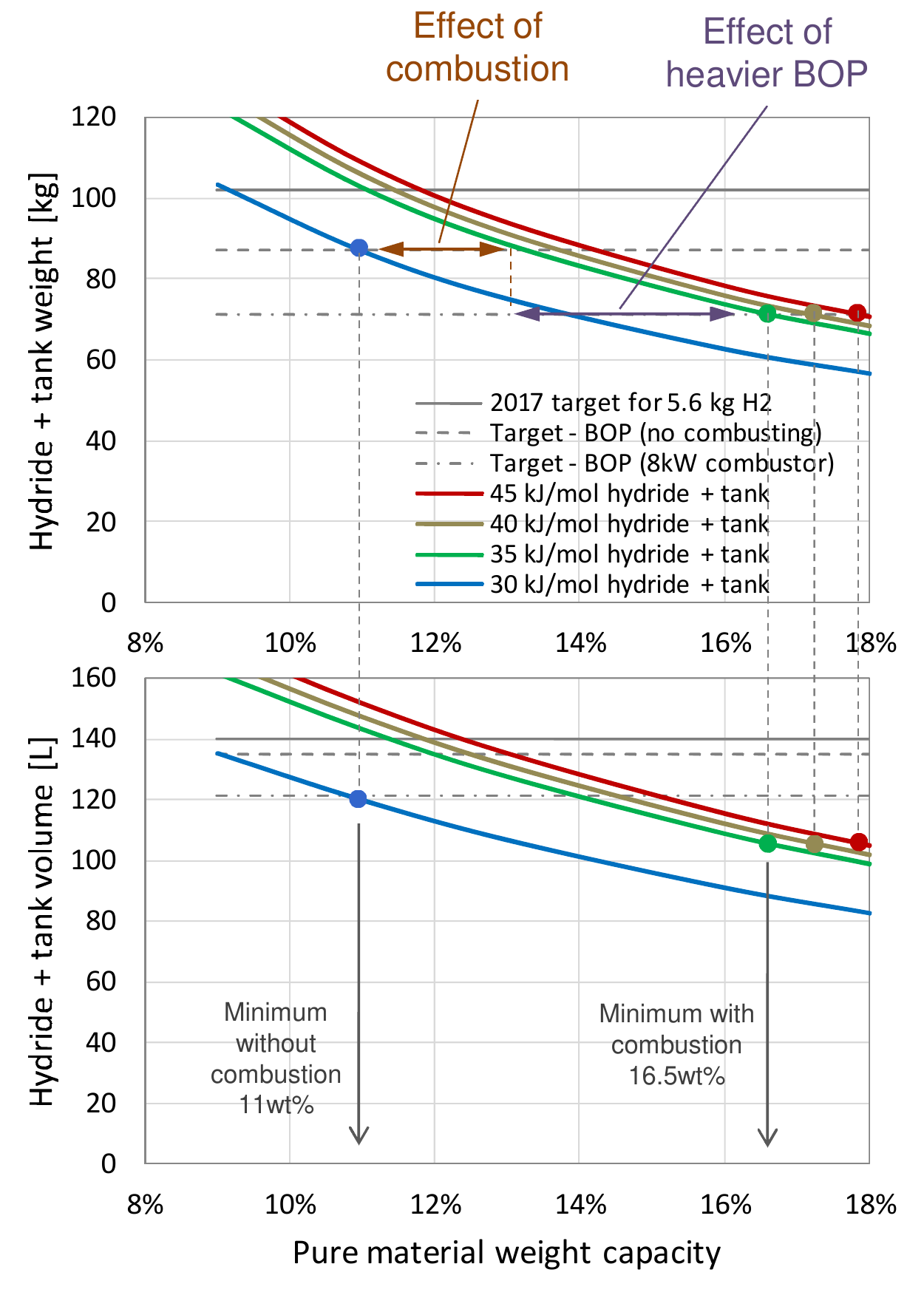}}%
\caption{Weight and volume of the pressure vessel and its internals as a function of the gravimetric capacity of the complex hydride pure material. The solid gray line is the maximum system weight and volume from the 2017 DOE targets. The dashed line is the maximum minus the minimum BOP for a system without a combustor (Fig.~\ref{minimal_system_no_combustion}). The dot-dashed line is the maximum minus the minimal BOP for a system with a combustor (Fig.~\ref{minimal_system_with_combustion}).}%
\label{tank_weight_vs_capacity}%
\end{figure}

Fig.~\ref{tank_weight_vs_capacity} shows the resulting weight and volume of the pressure vessel and all its internals, including the hydride, for complex metal hydrides. As the gravimetric capacity of the material increases, the tank becomes lighter and more compact. Part of the reduction is due to requiring less hydride and part is due to a lighter and smaller pressure vessel. The figure also shows the system weight and volume limits required by the 2017 DOE targets for a 5.6~kg-H$_2$ system, as well as the limits for systems with and without a combustor loop, using the numbers from Table~\ref{tab:bop}. Due to thermodynamics (see Fig.~\ref{equilibrium_pressure}), a 30~kJ/mol system could run entirely on the fuel cell waste heat. Therefore a material with that enthalpy and 11~wt\% capacity could be built into a system that satisfies the 2017 weight and volume targets. Furthermore, that system would automatically satisfy the 90\% on-board efficiency target.

As the enthalpy increases, there is first a dramatic jump in the curves between 30 and 35~kJ/mol due to the change in equilibrium pressure: to effect the dehydrogenation, the tank operating temperature set point must be increased beyond what the fuel cell can reliably provide as waste heat (see Fig.~\ref{equilibrium_pressure}). To compound the problem, this larger and heavier tank must be part of a system that also includes a combustor loop, and therefore its upper limit is given by the dot-dashed line in Fig.~\ref{tank_weight_vs_capacity}. Thus a 35~kJ/mol material would need 16.5~wt\% capacity to allow a system to be built around it while still satisfying the 2017 weight and volume targets. Referring back to Fig.~\ref{h2_combusted}, we see that if a 30~kJ/mol material has to operate at high enough temperature to require full combustion, the system automatically needs to combust at least $\sim$14\% of the hydrogen, and therefore cannot possibly satisfy the 90\% on-board efficiency target. For higher enthalpy materials the situation is even worse.

Note that for destabilized systems the situation is different. From Fig.~\ref{equilibrium_pressure} we see that such materials, because of their higher equilibrium pressure, start requiring combustion at lower enthalpy, with the transition being around $\sim$18~kJ/mol. This means that once combustion is required, the system can still satisfy the efficiency target, as long as the enthalpy is not increased too much.

\begin{figure}[tb]%
\centerline{%
\includegraphics[width=70mm]{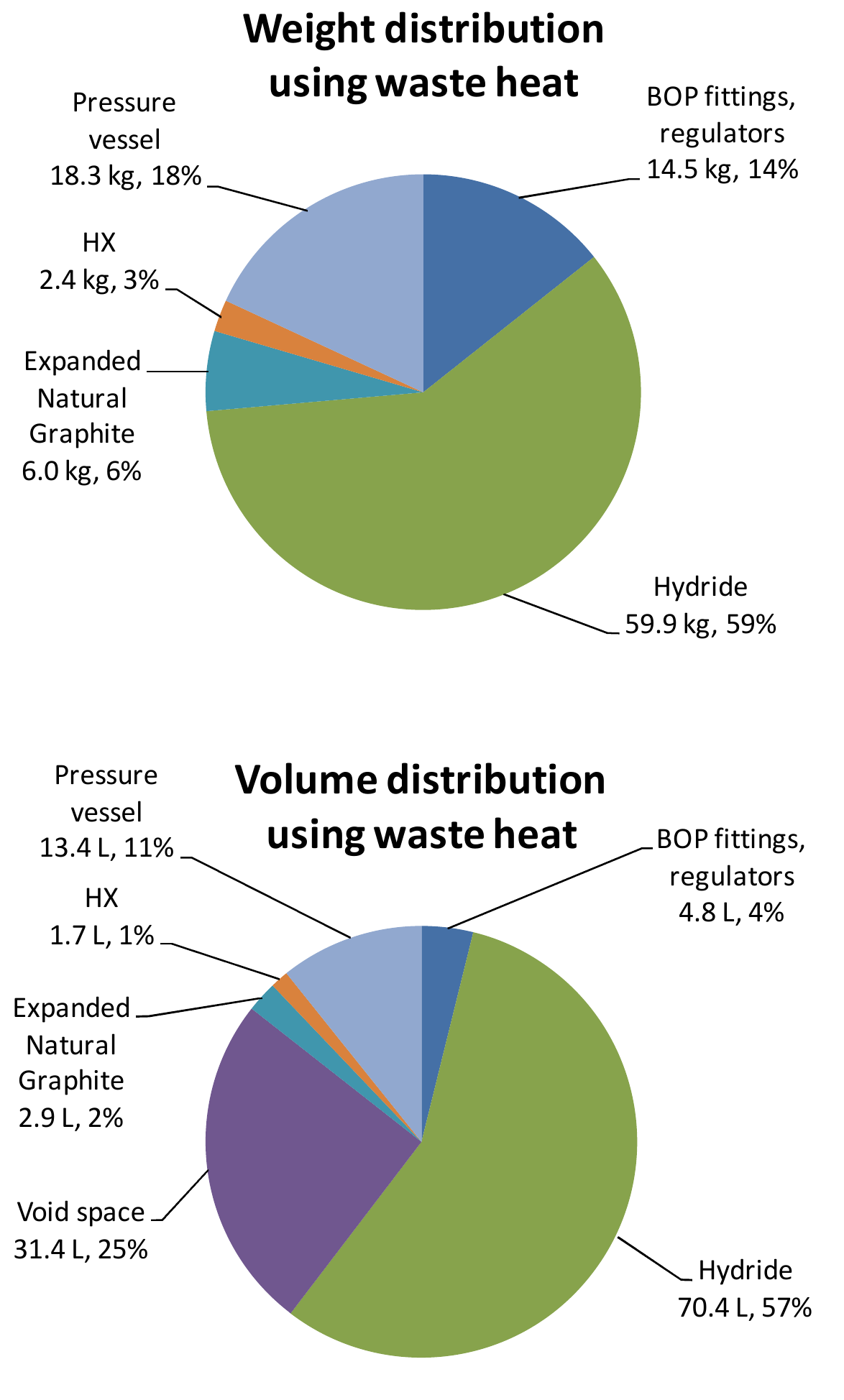}}%
\caption{Weight (kg) and volume (L) distribution for a system with 11 wt\% capacity (pure material) and $\Delta H = -27$~kJ/mol-H$_2$. This system can run entirely on the fuel cell waste heat.}%
\label{weight_and_volume_distribution_no_combustion}%
\end{figure}

\begin{figure}[tb]%
\centerline{%
\includegraphics[width=70mm]{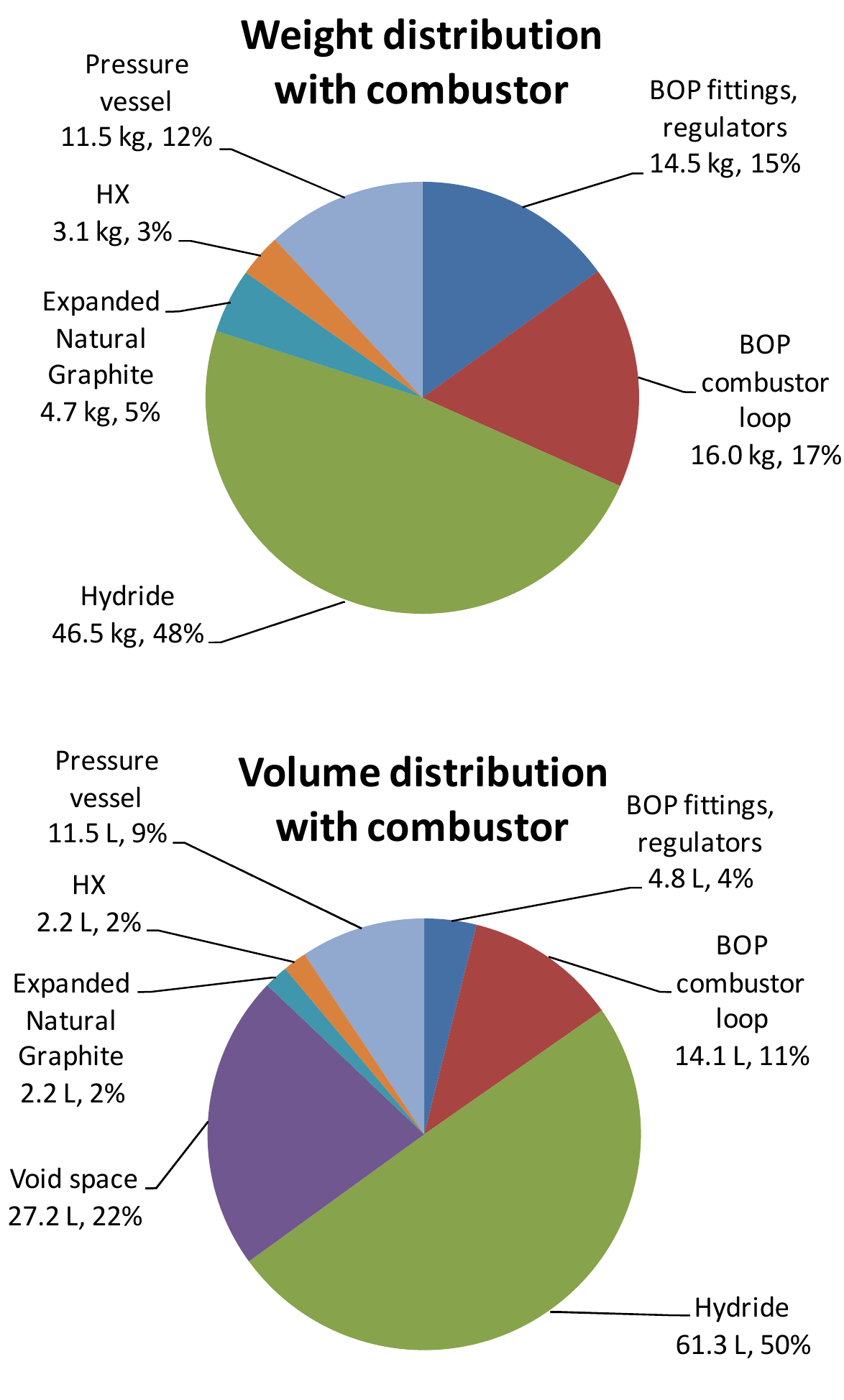}}%
\caption{Weight (kg) and volume (L) distribution for a system with 17 wt\% capacity (pure material) and $\Delta H = -40$~kJ/mol-H$_2$. This system requires a combustor and does not satisfy the on-board efficiency requirement.}%
\label{weight_and_volume_distribution_with_combustion}%
\end{figure}

Fig.~\ref{weight_and_volume_distribution_no_combustion} shows the weight and volume distribution for a system that does not require a combustor loop. Fig.~\ref{weight_and_volume_distribution_with_combustion}, on the other hand, shows the different weight and volume contributions to a system that requires combustion.

\subsection{Material requirements}

\begin{figure}[tb]%
\centerline{%
\includegraphics[angle=180,width=90mm]{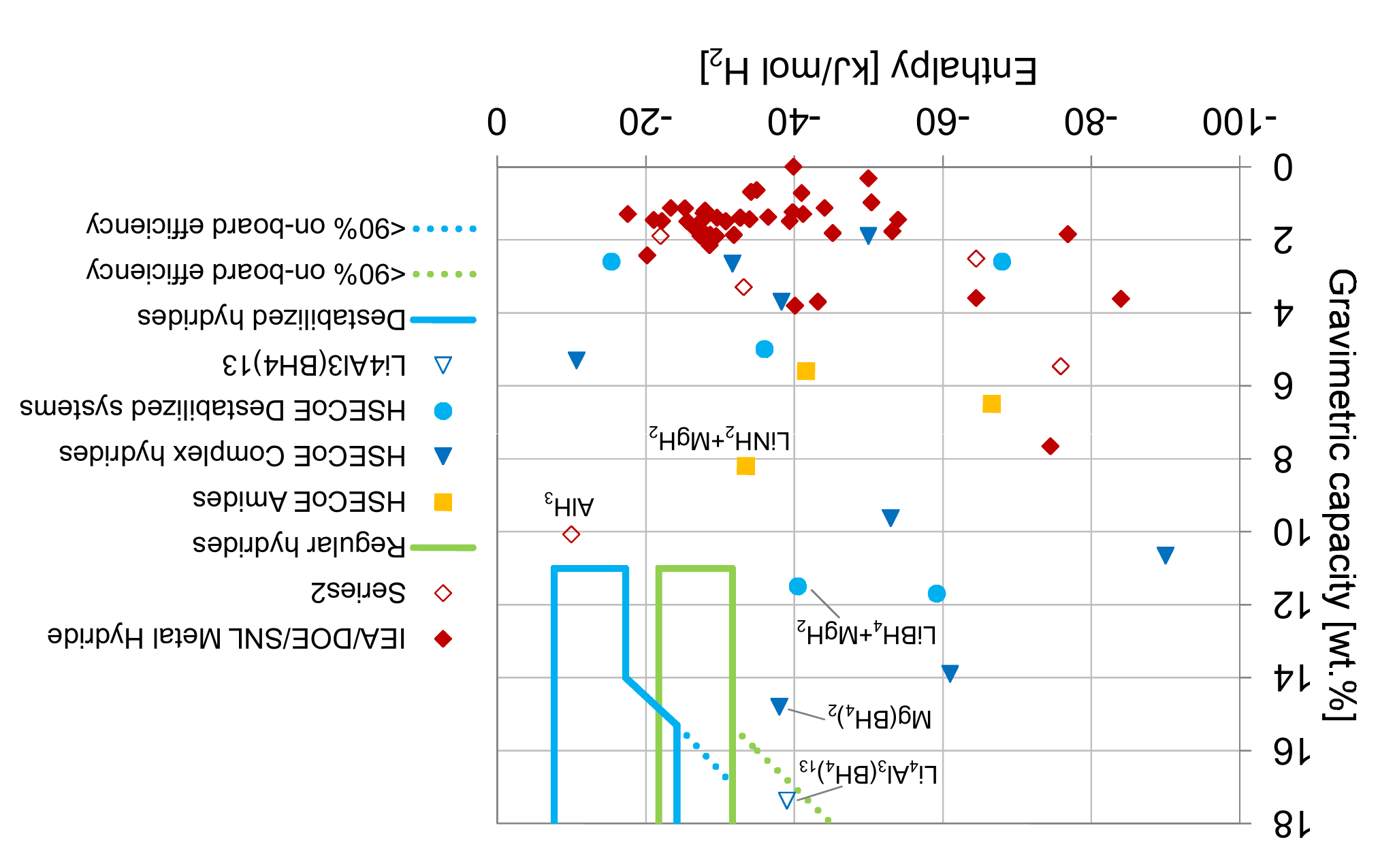}}%
\caption{Enthalpy and gravimetric capacity of metal hydrides. The green (resp. blue) window encloses the region where systems of regular (resp. destabilized) metal hydrides with minimal balance of plant components can be designed that satisfy the DOE 2017 targets.}%
\label{material_window}%
\end{figure}

Fig.~\ref{material_window} summarizes the material requirements on enthalpy and gravimetric capacity of the pure material. The solid green line encloses the window where systems of regular metal hydrides can be designed to satisfy the DOE 2017 targets. The solid blue line encloses the corresponding window for destabilized metal hydrides. The difference between the two families of materials is driven by the difference in the entropy fits in Fig.~\ref{entropy_vs_enthalpy}, which implies different equilibrium pressure for the same enthalpy. We now follow both windows clockwise, starting from the top-right. 

The lower enthalpy limit (rightmost vertical line) in each window is driven by thermodynamics. We assumed a pressure vessel rated for 100~bar, and that line corresponds to the enthalpy that yields 100~bar at 60$\degree$C. To the right of this line (lower enthalpy) the pressure is too high at ambient conditions that may be encountered. Continuing clockwise, the lower horizontal line (11~wt\%) represents the minimum gravimetric capacity where the system can accommodate enough hydrogen in the required short refueling time as well as the minimum BOP components when no combustion loop is needed.

The next corner represents the point where some hydrogen must be combusted in order to operate the system, and a combustion loop must be accommodated. This implies that the gravimetric capacity must increase accordingly. Here the two families of hydrides differ qualitatively. The diagonal segment represents the region where the system cannot run on fuel cell waste heat and must use some of the stored hydrogen for maintaining the release. As the enthalpy increases, more hydrogen must be stored. In order not to add to the system weight, this must be compensated with increased gravimetric capacity. As the fraction of combusted hydrogen increases, eventually DOE's target of 90\% on-board efficiency ceases to be satisfied. This is depicted as a dotted continuation of the solid diagonal line in Fig.~\ref{material_window}.

The region of interest in Fig.~\ref{material_window} is cornered by an interesting set of metal borohydrides (Li$_4$Al$_3$(BH$_4$)$_{13}$, Mg(BH$_4$)$_2$, and LiBH$_4$+MgH$_2$) with a high theoretical gravimetric capacity on the high enthalpy side and the LiNH$_2$--MgH$_2$ system on the low gravimetric density side. The higher than desired enthalpy of the metal borohydrides reflects their high thermal stability and reduces the on-board efficiency, as waste heat from conventional PEM fuel cells will no longer be sufficient to release the stored hydrogen. Except for the destabilized LiBH$_4$--MgH$_2$ composite system, on-board regeneration to full H$_2$ storage capacity at moderate temperature and hydrogen pressure has yet to be demonstrated. The LiNH$_2$--MgH$_2$ material brackets the region of interest at the bottom of the region of interest with a theoretical 8 wt\% gravimetric capacity, which is lower than desired, but the effect of the additional weight on the range of the vehicle is small. An important concern for all these high capacity materials is the rate of H$_2$ absorption and desorption. Substantial progress has yet to be made in order to achieve practical H$_2$ storage systems for light-duty vehicles for which the DOE has set a refueling time target of 3.3 minutes for 5~kg of usable H$_2$. Stability of the H$_2$ storage capacity is affected by the release of species that are considered contaminants for the PEM fuel cell such as NH$_3$ and B$_2$H$_6$, which will require a cleanup step.

We reiterate, however, that these windows in Fig.~\ref{material_window} were built from targets for one specific application (on-board rechargeable light-duty vehicles), and does not constitute a general statement for other applications and modes of use, such as those described in Ref.~\cite{McWhorter2011}, where many of the current hydrides have attractive and competitive characteristics.

\section{Summary and conclusions}

In this paper we combined simplified storage system and vehicle models with metal hydride databases in order to obtain material-level requirements for metal hydrides that can satisfy the DOE 2017 system-level targets.  The tank sizing models focus on the heat exchanger design to reject the heat of absorption during the short refueling time. We assume minimal BOP components for the rest of the system, considering both systems with and without a hydrogen combustion loop for supplemental heating of the tank during operation.

As a result, we require a minimum 11~wt\% gravimetric capacity of the pure material for the narrow enthalpy range in which no hydrogen combustion is necessary. For destabilized systems, we show that some enthalpy range that requires a combustion loop may yield feasible systems, albeit with at least 14~wt\% capacity. For regular hydrides, as soon as combustion is required, the target for 90\% on-board efficiency is automatically violated.

The resulting viability windows, as shown in Fig.~\ref{material_window}, are currently not populated by metal hydrides, although some hydrides are somewhat close to it, but with either too high enthalpy or too low gravimetric capacity. High enthalpy leads to high temperature of dehydrogenation and consequently inevitably lower efficiency, as the fuel cell waste heat cannot be utilized. Future development of higher temperature PEM fuel cells may mitigate this problem for the most promising materials. To accommodate the low enthalpy, lower capacity material, on the other hand, somewhat reduced range could be balanced with somewhat increased weight and volume for the storage system without sacrificing efficiency. We emphasize that these conclusions are \emph{for the specific application of light-duty vehicles} and this precludes neither their viability in a different setting nor the possibility of future material developments~\cite{McWhorter2011}.

\section*{Acknowledgments}

The authors thank all members of the DOE Hydrogen Storage Engineering Center of Excellence, in particular Norman Newhouse and John Makinson (Lincoln Composites) for sizing the pressure vessels, Vinod Narayanan and Kevin Drost (Oregon State University) for sizing the microchannel combustor/heat exchanger, and Bruce Hardy and Donald Anton (SRNL) as well as Michael Veenstra (Ford) for fruitful discussions and critical review. We also thank Robert Bowman for detailed helpful comments, and  Ned Stetson and Jesse Adams for their outstanding support.

This paper was prepared as an account of work supported by the United States Department of Energy under Contracts No.\ DE--FC36--09GO19006 (UTRC), DE--AC09--08SR22470 (SRNS), DE--FC36--09GO19003 (GM), and DE--AC06--76RLO1830 (PNNL). Neither the United States Government nor any agency thereof, nor any of their employees, makes any warranty, express or implied, or assumes any legal liability or responsibility for the accuracy, completeness, or usefulness of any information, apparatus, product, or process disclosed, or represents that its use would not infringe privately owned rights. Reference herein to any specific commercial product, process, or service by trade name, trademark, manufacturer, or otherwise, does not necessarily constitute or imply its endorsement, recommendation, or favoring by the United States Government or any agency thereof. The views and opinions of authors expressed herein do not necessarily state or reflect those of the United States Government or any agency thereof.

\appendix




\end{document}